\documentclass{article}

\usepackage[creativeai, final]{neurips_2025}

\usepackage[utf8]{inputenc} 
\usepackage[T1]{fontenc}    
\usepackage{hyperref}       
\usepackage{url}            
\usepackage{booktabs}       
\usepackage{subcaption}
\usepackage{amsfonts}       
\usepackage{nicefrac}       
\usepackage{microtype}      
\usepackage{xcolor}         
\usepackage{graphicx}
\usepackage{wrapfig}
\usepackage{soul}
\usepackage{tikz}
\usepackage{standalone}
\usetikzlibrary{calc,shapes.geometric}

\title{\textit{The Ghost in the Keys}: A Disklavier Demo for Human-AI Musical Co-Creativity} 

\author{
  Louis Bradshaw$^{1}$\thanks{Equal contribution} \quad Alexander Spangher$^{2}$\footnotemark[1] \quad Stella Biderman$^{3}$ \quad Simon Colton$^{1}$ \vspace{2mm}\\
  $^1$Queen Mary University of London\\
  $^2$Stanford University\\
  $^3$EleutherAI\\
  \texttt{louis.bradshaw@qmul.ac.uk}
}

\begin{document}

\maketitle

\begin{abstract}

\setcounter{footnote}{0}
While generative models for music composition are increasingly capable, their adoption by musicians is hindered by text-prompting, an asynchronous workflow disconnected from the embodied, responsive nature of instrumental performance. To address this, we introduce \textit{Aria-Duet}\footnote{Available at: \href{https://github.com/eleutherai/aria}{https://github.com/eleutherai/aria}}, an interactive system facilitating a real-time musical duet between a human pianist and \textit{Aria}, a state-of-the-art generative model, using a Yamaha Disklavier as a shared physical interface. The framework enables a turn-taking collaboration: the user performs, signals a handover, and the model generates a coherent continuation performed acoustically on the piano. Beyond describing the technical architecture enabling this low-latency interaction, we analyze the system's output from a musicological perspective, finding the model can maintain stylistic semantics and develop coherent phrasal ideas, demonstrating that such embodied systems can engage in musically sophisticated dialogue and open a promising new path for human-AI co-creation. 

\end{abstract}

\section{Introduction}

The adoption of modern AI music tools suggests a notable trend. While tools for passive tasks like source separation appear to be widely adopted into the creative workflows of producers, models that generate core compositional content have seemingly seen slower uptake among classically trained musicians and composers. This perceived gap is not, we argue, because musicians are averse to ceding creative control; it is caused, rather, by the \textit{mode of interaction} presented by current AI models for musical composition. The paradigm of asynchronous text-prompting and speed-bottlenecked iteration is fundamentally at odds with the embodied, responsive, and often non-verbal feedback loops that define social \textit{musicking} \citep{small1998musicking} and \textit{creative flow} \citep{csikszentmihalyi1990flow}.

Aiming to bridge this gap for pianist-composers, we introduce \textit{Aria-Duet}, an interactive system for real-time musical duets between a pianist and generative model. The system's physical interface is a Yamaha Disklavier, an acoustic piano capable of both capturing and physically playing performances through a MIDI connection. A musician plays on the instrument while the model listens; then, upon a \textit{takeover signal}, the system responds by generating and playing a continuation on the same keys in real-time. This interaction is powered by \textit{Aria} \citep{bradshaw2025scaling}, an autoregressive transformer \citep{vaswani2017attention, huang2018music} trained to compose expressive piano continuations one note at a time.

\textit{Aria-Duet} is designed to recenter the artist's creative agency by restoring familiar feedback loops and enabling \textit{experimental play}. This interaction is facilitated by the model's ability to adapt to a broad range of musical styles, vocabularies, and forms. However, realizing an experience that feels genuinely fluid and engaging is not merely a matter of connecting a model to an instrument. As we will detail, the success of this interaction hinges on addressing key design challenges, including minimizing takeover latency, ensuring musical coherence, and maintaining accurate acoustic playback.

We conclude by demonstrating and providing a musicological analysis of the system's output. By examining our system's responses to varied prompts, we highlight its ability to maintain stylistic semantics, develop coherent phrasal ideas, and even engage in multi-voice dialogue. Overall, our explorations contribute toward a practical blueprint for designing real-time, interactive AI systems that augment the human creative process.

\section{Related Work}

Our work builds on decades of research into human-computer musical co-creation. Early interactive systems, such as George Lewis's \textit{Voyager} \citep{lewis1999voyager} and Biles's \textit{GenJam} \citep{biles1994genjam} pioneered real-time interaction using rule-based grammars or genetic algorithms. While inventive, these systems were often limited to pre-programmed styles and struggled to adapt to unexpected musical inputs from the performer.

The subsequent wave of \textit{statistical style-learning} companions enabled systems to learn directly from a performer's input. The most influential of these, Pachet's \textit{Continuator} \citep{pachet2003continuator}, introduced a back-and-forth keyboard partnership using Markov models that directly inspires our system's interactive paradigm. However, like other systems of its era (e.g., \textit{BoB} \citep{thom2000bob}, \textit{OMax} \citep{assayag2006omax}), its reliance on methods with short context windows resulted in compositions which often failed to capture the deeper phrasal and structural logic typical of human-composed music.

The deep learning era brought architectures capable of modeling long-range dependencies. Systems like Google's \textit{AI Duet} (using LSTMs \citep{eck2017aiduet}) and later Transformer-based models like \textit{Music Transformer} \citep{huang2018music} and \textit{MuseNet} \citep{openai2019musenet} demonstrated the ability to generate multi-bar structures and richer harmonic vocabularies. Recent work on interactive systems, such as \textit{ReaLJam} \citep{scarlatos2025realjamrealtimehumanaimusic} and \textit{Jam Bot} \citep{blanchard2025jam}, have explored applications of modern neural models in real-time contexts. However, a gap remains: systems often sidestep the critical engineering challenges that are essential for a truly fluid, embodied interaction through acoustic instruments (e.g., Disklavier pianos). \textit{Aria-Duet} directly addresses this gap by integrating a state-of-the-art model with a system meticulously designed for low-latency, acoustic performance.

\section{System Design}

\textit{Aria-Duet} is comprised of two primary components: (1) a generative model for piano performance, adapted from \textit{Aria} \citep{bradshaw2025scaling}, used to generate creative and expressive continuations of a user's performance on the Disklavier, and (2) a real-time engine that manages the user control flow, input/output, and real-time inference. In this section, we outline the design and implementation of each component.

\subsection{Generative Model}

Our system's continuation is generated by a model finetuned from \textit{Aria} \citep{bradshaw2025scaling}, an autoregressive transformer model designed to model expressive symbolic piano performances (i.e., on the note-level). \textit{Aria} is particularly well-suited for this application due to its training and tokenization scheme. It was pretrained on a refined subset of \textit{Aria-MIDI} \citep{bradshawaria}, a large-scale (100k+ hours) dataset of solo piano music spanning a wide range of genres and styles. This dataset was curated using a transcription model that was itself trained on paired audio and MIDI recordings from a Disklavier \citep{hawthorne2018enabling}. This creates a direct correspondence between the model's training data and the Disklavier-based I/O of our system. \textit{Aria} employs a note-centric tokenizer that quantizes musical events with a fine-grained resolution, generating continuations by performing next-token (i.e., next-note) prediction in an iterative process.

In a preliminary version of our system, we used the native pretrained \textit{Aria} model for generation. However, informal testing with pianists revealed two shortcomings that detracted from the co-creative experience. First, the model lacked explicit sustain pedal tokens, instead simulating sustain with immediate note retriggerings. While functionally equivalent in software, pianists universally found this approach jarring when played back on the Disklavier, which requires a gap to retrigger notes. One pianist also noted that this simulation only sustained notes, missing the acoustic resonance created by lifting the dampers. Second, due to the overrepresentation of popular works in the original training data, the model would often complete a famous theme when prompted. Classically trained pianists, in particular, found this frustrating, as it disrupted their natural tendency to use well-known repertoire as a starting point. To address these issues, we post-trained the model on a high-quality deduplicated subset of the \textit{Aria-MIDI} dataset that included explicit pedal-on and pedal-off tokens. This single intervention addressed both problems, enabling the system to control the pedal while mitigating its tendency toward compositional memorization.

\subsection{Real-time Engine}

\begin{figure}[t]
\includegraphics[width=\linewidth]{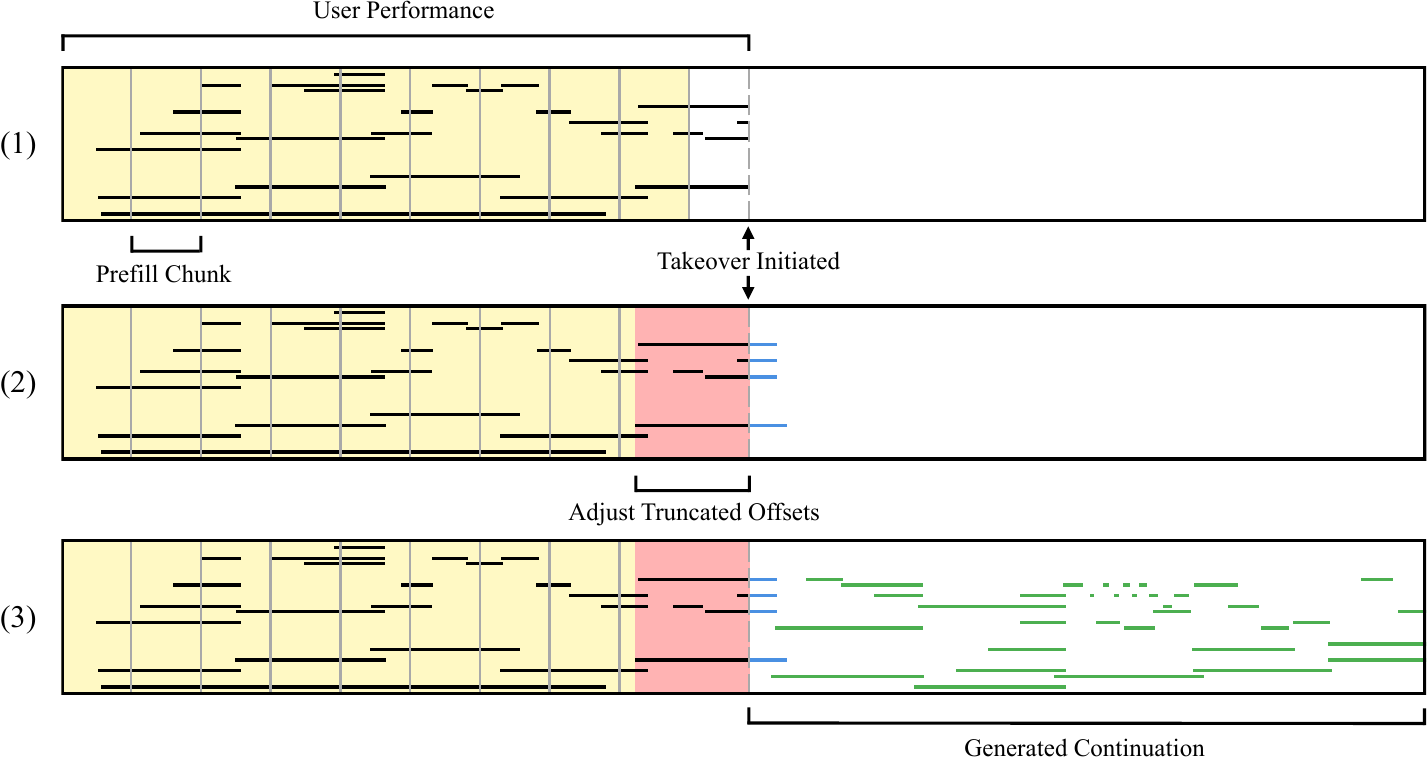}
  \caption{An illustration of the KV-Cache management for real-time, low-latency operation. (1) \textit{Listen}: As the user plays, the received context is proactively and continuously prefilled into the model's KV-Cache in chunks. (2) \textit{Takeover}: Upon a takeover signal, the system finalizes the input, prefilling any missing context and speculatively re-evaluating the durations of any hanging notes (seen in blue), ensuring a seamless transition and preparing the KV-Cache. (3) \textit{Generate}: The model then begins generating a musical continuation note-by-note, streaming the result to the Disklavier.}
\label{fig:kv-cache}
\end{figure}

Our system's design is built upon an embodied, turn-based interaction model facilitated entirely by a Disklavier piano. The operational flow is as follows: a pianist begins by playing, and their performance is captured and routed to a computer as a stream of MIDI. To cede control to the generative model, the pianist presses the left pedal (\textit{una corda}), which serves as the takeover signal. This triggers the system, running on a connected Apple Silicon device, to pre-process the performance and prompt the model to generate a musical continuation, which is streamed back to the Disklavier and performed on the keys in real-time. The pianist can reclaim control at any point by re-pressing the pedal, enabling an alternating musical dialogue that preserves the musical context. This control scheme aims to be intuitive, mapping system-level commands to familiar physical actions; however, to keep the interaction frictionless, several key engineering challenges must be addressed.

\paragraph{Response Latency.}
The inference process for autoregressive transformers involves two distinct computational phases: \textit{prefill} and \textit{decoding}. During prefill, the model processes the entire input prompt in parallel, a compute-bound (FLOPs) operation that populates a key-value (KV) cache with the context's attention states. Subsequently, the decoding phase generates the output one token at a time in an iterative process that is memory-bandwidth-bound. This presents a key challenge for real-time interaction on consumer hardware like Apple Silicon. While this hardware's high memory bandwidth is well-suited for fast decoding, the compute-bound prefill phase creates a significant bottleneck, introducing an unacceptable lag of 1000-2000ms between the user's takeover signal and the model's first note. To mitigate this latency, we implement a \textit{continuous prefill} strategy that proactively updates the KV-cache in small chunks as the user plays. This distributes the computational load over time, virtually eliminating the prefill-induced delay at the moment of transition.

\paragraph{Transition Coherence.}
Beyond minimizing the \textit{time-to-first-note}, the musical coherence of the transition is equally important to the user experience. The core challenge stems from the fact that the model, trained to continue any input, is highly sensitive to the musical context immediately preceding the takeover. A common scenario where this manifests is when the user initiates a transition while notes are still held or sustained by the pedal. Due to the tokenizer's design, the model must be provided with complete note information, including durations, before it can predict new notes. A naive approach would be to force-end all active notes at the transition point. This, however, provides a 'truncated' context that often corrupts the model's predictions, causing it to generate similarly abrupt, staccato phrases. To remedy this, our system speculatively reevaluates the durations of notes truncated by the transition, filling these corrected durations into the model's KV-cache before generating a continuation. While this process adds a small latency overhead of \textasciitilde100-200ms, it is vital for ensuring a smooth and musically coherent transition. Figure \ref{fig:kv-cache} illustrates the state of the KV-cache and how it is modified during these phases of operation.

\paragraph{Disklavier Playback.}
Translating the model's generated output into an accurate performance requires addressing the physical limitations of the Disklavier. Unlike for software synthesizers, the electromechanical action of a piano introduces two challenges: velocity-dependent note-on latency, where louder notes sound with different delays than softer ones; and mechanical conflicts, where a key cannot be retriggered before its action has physically reset. The Disklavier’s native \textit{playback mode} resolves these issues by using a buffer, but at the cost of a fixed 500ms latency that detracts from interactive use. Our system circumvents this by implementing a custom, zero-latency streaming layer that modifies the playback schedule in real-time. Instead of buffering, it makes two just-in-time adjustments: First, it schedules the send-time for each note-on message to account for manually-calibrated velocity-specific latency, and only articulates notes whose scheduled time arrives before exceeding a staleness threshold. Second, to prevent re-articulation errors, the system retrospectively modifies the send-time of a pending note-off message if a new note-on for the same pitch is generated before the off-message has been sent. This dynamic rescheduling enforces the necessary physical gap between notes by altering the timing of a future event, rather than by introducing a processing delay. 

\section{Demonstration}

\begin{wrapfigure}[15]{l}{0.45\linewidth}
    \vspace{-10pt} 
    \centering
    \begin{subfigure}{0.48\linewidth}
        \includegraphics[width=\linewidth, height=1.35\linewidth]{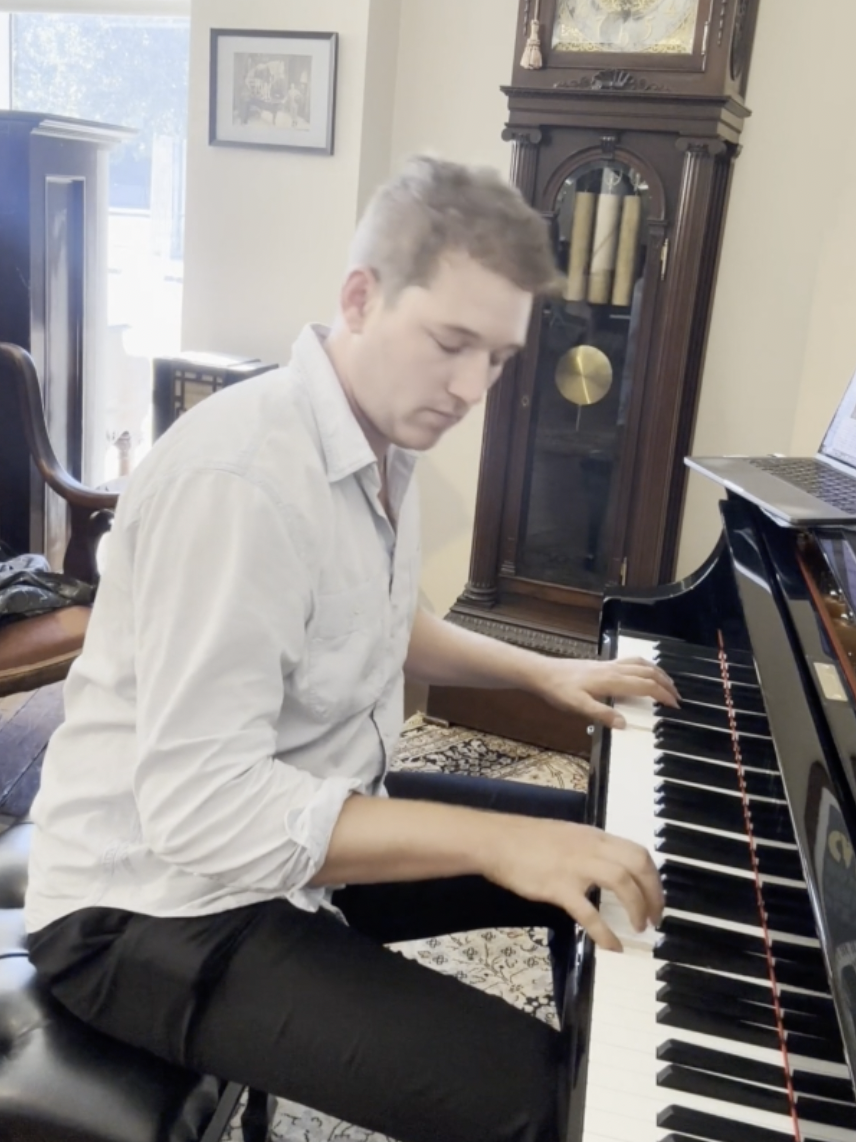}
        \caption{Musician prompts \textit{Aria-Duet}.}
    \end{subfigure}%
    \hfill
    \begin{subfigure}{0.48\linewidth}
        \includegraphics[width=\linewidth, height=1.35\linewidth]{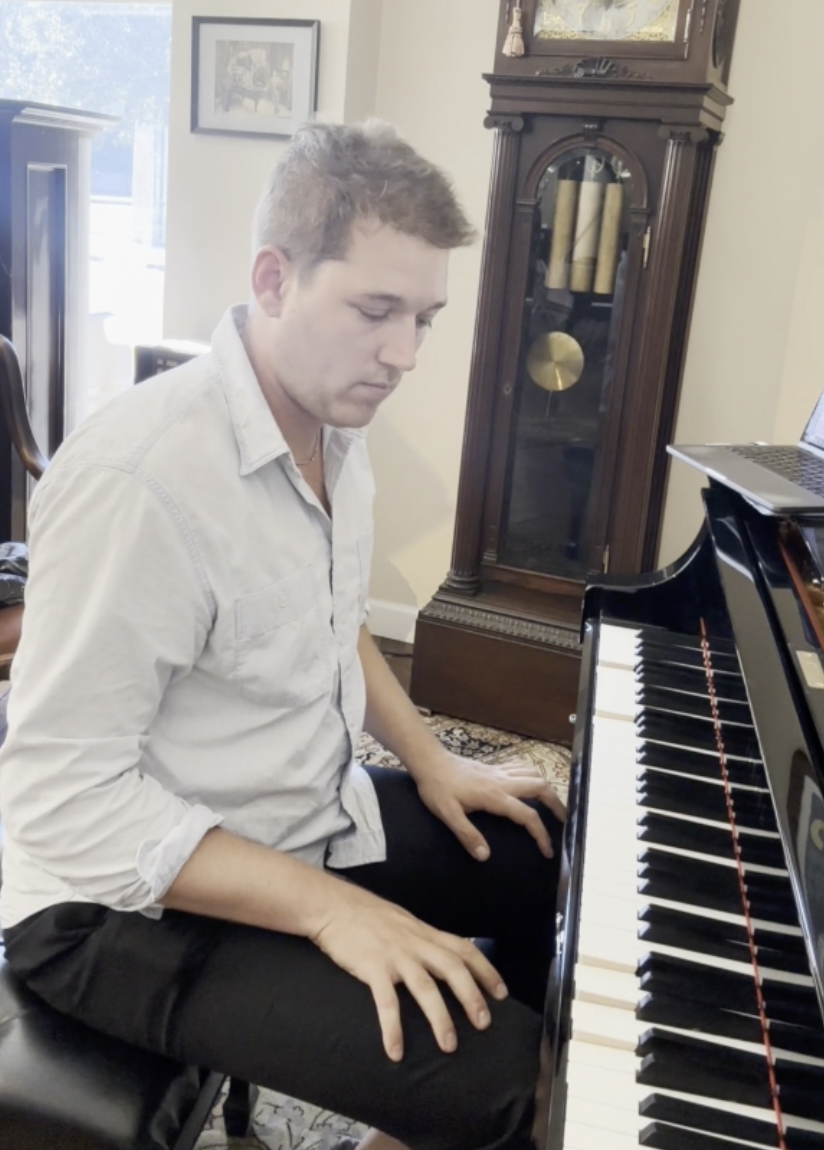}
        \caption{\textit{Aria-Duet} produces a response.}
    \end{subfigure}
    \caption{Musician interacts with the \textit{Aria-Duet} system using a Disklavier.}
    \vspace{-20pt}
\end{wrapfigure}

In this section, we analyze the musicality of Aria-Duet's generated performances. Using a three-minute video demonstration\footnote{Available at: \href{https://youtu.be/8s3V922h3CU}{https://youtu.be/8s3V922h3CU}}, we explore how the system responds to a variety of prompts from a human pianist. In the demonstration are prompts to the following songs: \textit{Carmen Overture} (Georges Bizet, 1875), \textit{Ave Maria} (Franz Schubert, 1825; arranged by Franz Liszt, 1846), \textit{Amazing Grace} (William Walker, 1835), \textit{The Entertainer} (Scott Joplin, 1902), \textit{Freygish Fino} (Noah Smith, 2024), \textit{Spanish Song} (original composition, 2025), \textit{Piano Concerto No. 2} (Sergei Rachmaninoff, 1901), and \textit{Hungarian Rhapsody No. 6} (Franz Liszt, 1847).

\paragraph{\textit{Semantics}: Vocabularies and Prosody.} According to \citet{langer1942philosophy}, \textit{vocabularies} emerge across modalities to meet the expressive needs of social cultures. \textit{Musical} vocabularies emerge \textit{both within songs and across genres} as \textit{small motifs and patterns emerge and vary} \citep{schoenberg1984style, narmour1992analysis}. Maintaining a piece's musical vocabulary is crucial for maintaining its coherence for the listener, thus, we explore how well \textit{Aria} can maintain distributional semantics in given prompts. In our demonstration, we examine how \textit{Aria} responds to prompts using less common tonalities, rhythms and chordal progressions\footnote{Less common with respect to the Western classical canon, which constitutes the bulk of \textit{Aria's} training data \citep{bradshawaria}, \citep{bradshaw2025scaling}.}: for \textit{Aria} to demonstrate a deep understanding of vocabulary, it must generate continuations adhering to these semantics and not simply revert to modal norms. 

\textit{Spanish Song} and \textit{Freygish Fino} make use of the Freygish (Phrygian dominant) and Phrygian scales, which are relatively uncommon in mainstream Western tonal music compared to the diatonic major or minor scales. As can be seen in both instances, Aria is naturally able to identify and improvise within the appropriate scales; in \textit{Freygish Fino}, it even continues and expands on the \textit{Andalusian cadence}, a traditional Klezmer chord progression. \textit{The Entertainer} offers an additional example of semantic consistency: \textit{The Entertainer} uses the ragtime rhythmic structure, which consists of syncopation patterns and emphases on the 2/4 beats -- both of which are outside standard Western classical music. \textit{Aria} continues to maintain this pattern across an entire musical phrase, keeping the player and listener in the style of the prompt. 

We further examine another dimension of style called \textit{musical prosody}. \textit{Prosody}, in spoken language, describes the way vocal inflections change the meaning of utterances \citep{shriberg1998can} (e.g., ``\textit{Hellllllo}!'' spoken brightly conveys a different meaning than a stern, curt ``\textit{hello.}''), despite mapping to the same vocabulary token. In music, performance dynamics and micro-improvisations on the melodic line (e.g., playing a phrase loudly, faster, slower, etc.) can have a similar prosodic effect \citep{heffner2015prosodic}. In this demo, \textit{Amazing Grace} offers an example of musical prosody: the player improvises micro-elements, like suspensions and rhythmic variations. Surprisingly, Aria \textit{follows} these unseen variations, creating a novel continuation with prosodic consistency.

\paragraph{\textit{Dialogue}: Harmony and Counterpoint.}
Harmony and counterpoint are two ways a song weaves multiple different lines in real-time \citep{prout1891fugue, Piston1947} and are seen as playing a key role in a listener's musical comprehension, by connecting the musical act to multi-speaker dialogues \citep{Cone1974, johnson1986analytical, Klorman2016}. For \textit{Aria} to maintain multiple voices, it is crucial for it to generate comprehensible, nuanced outputs. Liszt's \textit{Ave Maria} transcription is notable for its multi-voice complexity: it maintains three stylistically distinct, coherent voices that the piano player weaves together: the soprano line, with far-reaching chordal leaps, the alto carrying the main melody, and the bass with a staccato grounding. \textit{Aria} continues these three voices and, further, captures the essence of each voice in its variations: each voice continues its distinctive character and role throughout the demonstration.

\paragraph{\textit{Discourse}: Phrases and Structure}
Musical structure, shown both in phrasal consistency and macro-structural adherence, is an even higher-level component of sense-making \citep{schoenberg1999fundamentals, rosner1984generative}. \citet{rosner1984generative} explores \textit{phrasal structure} in songs, drawing inspiration from linguistic parse trees developed for clause-and sentence-level analysis. Much in the same way a linguistic clause captures a single verbal idea, a musical phrase captures a single musical idea. For a song to comprehensibly express ideas to listeners, it must \textit{introduce} and \textit{develop} phrases. Rachmaninoff's \textit{Piano Concerto No. 2} demonstrates  \textit{Aria}'s \textit{phrasal} sophistication. The prompt contains a phrase taken from the first movement's secondary theme \citep{yang2025musical}, featuring a musical arc -- i.e., the primary melody weaves up and down, along with a counterpoint bass line. \textit{Aria} develops this phrase, repeating the same arc several times through different keys, and maintaining the bass line. The idea develops without losing consistency. 

According to \citet{schoenberg1999fundamentals}, a song's macro-structure or form allows audiences to understand musical phrases within broader contexts and make sense of narrative arcs. In writing, macro-structure is often studied in the context of discourse structure  (e.g., essays \citep{Montaigne1580}, news \citep{van1988news}), where structural elements have semantic meaning (e.g., \textit{Introduction}, \textit{Thesis}). In music, macro-structure can also be observed in two ways: (1) \textit{semantic} awareness, or through elements which have meaning on their own (e.g., an \textit{Overture} introduces, a \textit{Coda} closes) \citep{Caplin1998}; or (2) \textit{formal structural repetition}, or elements which take on meaning because they occur and reoccur intentionally (e.g., $A\rightarrow B \rightarrow A$ progressions, where the A section reoccurs after the B section) \citep{schoenberg1999fundamentals}. Generative models have been observed to struggle with structural coherence \citep{bhandari2024motifs, spangher2022sequentially, sanchez2024stay} in some domains, although others have noted that different training approaches can induce more structural adherence \citep{tian2024large}.

Carmen's \textit{Overture} and Liszt's \textit{Hungarian Rhapsody No. 6} provide us with two examples of \textit{Aria's} semantic awareness: the \textit{Overture} opens the demonstration while the \textit{Rhapsody} closes it. In each, \textit{Aria} immediately recognizes the narrative requirements posed by the prompt, and generates continuations that fulfill the narrative purpose. In the first case, \textit{Aria} generates a slowly accelerating opening melody; in the second, it closes and ends the piece. \textit{Spanish Song} provides us an extended example of \textit{Aria's} formal structural awareness. The prompt presents a coherent melody, which \textit{Aria} repeats several times in its generation. It then goes on an extended interlude before bringing the melody back, at the end of the clip. Together, the prompt and completion represent a compelling AABA structure, where the B section provides a coherent bridge between the A sections.

\section{Discussion and Conclusion}

We are interested in whether \textit{Aria} is \textit{actually} displaying these aspects of musical comprehension, or whether it is simply memorizing examples from the training dataset. While this is interesting from a scientific perspective, it also matters practically for users. Researchers have explored the effects of both \textit{generating} novel responses and \textit{retrieving} existing data on creative workflows \citep{VanGenugten2022, Chavula2023, GindertMuller2024}; and both have been shown to have a stimulating effect. However, \textit{generative} approaches have been shown to aid in creativity more than \textit{retrieval}-based approaches \citep{GindertMuller2024, Gu2025}. 

In none of the completions we examined do we find evidence of memorization. Even for well-known songs, \textit{Aria} is inventive and generates completions that, to our knowledge, are novel. To confirm that \textit{Aria} can respond to novel input, we include in our analysis two novel compositions: the \textit{Spanish} song written by one of the authors in 2023, and \textit{Freygish Fino} written by Noah Smith in 2024, a professor at the University of Washington. Neither song was included in the training data. While \textit{Aria} \textit{does} seem to be less inventive in these cases, mainly playing what sounds to be improvised cadenzas, it maintains all levels of coherence that we sought to examine.

While promising, more work is required to broadly assess the impact of the \textit{Aria-Duet} system on creativity and co-creation. Coherence, deeper musical understanding, and true generative abilities, which we have sought to explore in this work, are important characteristics for a creative assistant. So far, early results appear to be positive. The pianists we worked with were excited and enjoyed using the \textit{Aria-Duet} system: the continuations generated were interesting and the mode of interaction was stimulating. 

We acknowledge the risks of our system. Research is mixed and sharply polarized on whether modern AI-creativity tools help to stimulate and support human users. While some research points to AI tools enhancing creativity both on an individual \citep{DoshiHauser2024} and group level \citep{Holzner2025, MeinckeGirotra2024, LeeChung2024}, other studies question their capabilities \citep{spangher2024llms}, with some even showing harms \citep{Meincke2025, Kosmyna2025}. Music -- despite being a fundamentally social act \citep{small1998musicking} -- faces a crisis in musician loneliness and competition \citep{cara2022understanding}. While AI cannot supplant human social interaction, we hypothesize that co-creation systems have the potential to help build user skill and confidence, preserve musical forms and styles and induce greater musical participation and creation in the future. This work represents a first step in that direction, by bringing a state-of-the-art music model into a functional, usable format and demonstrating its efficacy on a Disklavier.

\newpage
\bibliographystyle{plainnat}
\bibliography{references}

\end{document}